\documentclass[
preprint,
prd,tightenlines,showpacs,nofootinbib,
eqsecnum,
amsfonts,amsmath,amssymb]{revtex4}

\usepackage{bm}
\usepackage{graphicx} 
\usepackage{hyperref}
\usepackage{mathrsfs}

\newcommand{\ud}{\mathrm{d}}
\newcommand{\uD}{\mathrm{D}}
\newcommand{\calW}{\mathcal{W}}

\allowdisplaybreaks

\begin{document}

\title{Dipolar Dark Matter and Dark Energy}

\author{Luc \textsc{Blanchet}}\email{blanchet@iap.fr}
\affiliation{$\mathcal{G}\mathbb{R}\varepsilon{\mathbb{C}}\mathcal{O}$ Institut d'Astrophysique de Paris --- UMR 7095 du CNRS, \\ Universit\'e Pierre \& Marie Curie, 98\textsuperscript{bis} boulevard Arago, 75014 Paris, France}

\author{Alexandre \textsc{Le Tiec}}\email{letiec@iap.fr}
\affiliation{$\mathcal{G}\mathbb{R}\varepsilon{\mathbb{C}}\mathcal{O}$ Institut d'Astrophysique de Paris --- UMR 7095 du CNRS, \\ Universit\'e Pierre \& Marie Curie, 98\textsuperscript{bis} boulevard Arago, 75014 Paris, France}

\date{\today}

\begin{abstract}
In previous work [L.~Blanchet and A.~Le~Tiec, Phys. Rev. D \textbf{78}, 024031 (2008)], a model of dark matter and dark energy based on the concept of gravitational polarization was investigated. This model was shown to recover the concordance cosmological scenario ($\Lambda$-CDM) at cosmological scales, and the phenomenology of the modified Newtonian dynamics (MOND) at galactic scales. In this article we prove that the model can be formulated with a simple and physically meaningful matter action in general relativity. We also provide alternative derivations of the main results of the model, and some details on the variation of the action.
\end{abstract}


\maketitle

\section{Introduction and motivation}
\label{secI}

The concordance cosmological model $\Lambda$-CDM brilliantly accounts for the precise measurements of the anisotropies in the cosmic microwave background (CMB) \cite{HuDo.02}, for the formation and growth of large scale structure as seen in deep redshift \cite{Bl.al.03} and weak lensing \cite{BaSc.01} surveys, and for the fainting of the light curves of very distant supernovae \cite{Ri.al.98,Pe.al.99}. The paramount conclusion is that the total mass-energy content of the Universe is made by $\Omega_\text{b}\simeq 4\%$ of ordinary (essentially baryonic) matter, $\Omega_\text{dm}\simeq 23\%$ of cold dark matter (CDM), and $\Omega_\text{de}\simeq 73\%$ of dark energy which could be in the form of a cosmological constant $\Lambda$. However, no experimental claim of direct discovery of a CDM particle has been confirmed, and the attempts at interpreting $\Lambda$ in terms of fundamental quantum mechanics have failed.

With the advent of high precision cosmic $N$-body simulations (see \cite{Be.98} for a review), the cosmological model has been extrapolated to the smaller scale of galactic systems, and suggests the existence of a specific CDM density profile around galaxies \cite{Na.al.97}. However, the simulated CDM halos face severe challenges when compared to observations. Most problematic is the generic formation of cusps of dark matter in the central regions of galaxies, while the rotation curves seem to favor a constant density profile in the core \cite{SaBu.00,Ge.al2.07}. We mention also the prediction of numerous but unseen satellites of large galaxies \cite{Mo.al.99,Kr.al.04}, and the recent evidence \cite{Bo.al.07} that tidal dwarf galaxies are dominated by dark matter --- a fact which is at odds with the CDM tenets \cite{Ge.al.07}. Furthermore, the most important challenge is that the CDM scenario falls short in explaining in a natural way Milgrom's law \cite{Mi1.83,Mi2.83,Mi3.83}, namely that the need for dark matter arises only in regions where the typical acceleration of ordinary matter (or, equivalently, the typical value of the gravitational field) is below some universal constant acceleration scale $a_0\simeq 1.2 \times 10^{-10}~\text{m}/\text{s}^2$. This law manifests itself particularly in the flat rotation curves of spiral galaxies, and in the baryonic Tully-Fisher relation. No convincing mechanism for incorporating an acceleration scale such as $a_0$ in the $N$-body simulated CDM halos has been found. Although it is possible that some of these problems will be solved within the CDM approach \cite{Sw.al.03,Sp.al.05}, it is very important to consider alternative solutions.

The most successful alternative approach to the problem of dark matter in galactic halos is MOND --- Milgrom's modified Newtonian dynamics \cite{Mi1.83,Mi2.83,Mi3.83}, which insists that there is no dark matter and we instead witness a violation of the Newtonian law of gravity. In MOND the true gravitational field $\bm{g}$ experienced by ordinary matter (stars and gas) differs from the Newtonian one, and obeys the modified Poisson equation \cite{BeMi.84}
\begin{equation}\label{mond}
	\bm{\nabla} \! \cdot \left( \mu \, \bm{g} \right) = -4 \pi \, \rho_\text{b} \, .
\end{equation}
We use bold-face notation to represent ordinary three-dimensional vectors and pose $G = 1$. Here $\rho_\text{b}$ is the density of baryonic matter, and $\mu$ is the MOND function which depends on the norm $g=|\bm{g}|$ of the gravitational field. In the regime of weak gravitational fields, $g\ll a_0$, we have $\mu(g) = g / a_0 + \mathcal{O}(g^2)$, while $\mu(g) \rightarrow 1$ when $g \gg a_0$, so as to recover the usual Poisson equation. Various forms of the interpolating function $\mu$ have been proposed to fit observations in the best way \cite{FaBi.05,SaNo.07}.

The ability of the formula \eqref{mond} to reproduce a wide variety of phenomena associated with dark matter halos is tremendous (see e.g. \cite{SaMc.02,TiCo.07}). However, because \eqref{mond} is non-relativistic, it does not allow one to answer questions related to cosmology. In particular, it is a great challenge to find a theory reproducing both MOND at galactic scales and $\Lambda$-CDM at cosmological scales. A number of relativistic field theories have been proposed, recovering \eqref{mond} in the non-relativistic limit, and sharing with MOND the idea that dark matter is an apparent reflection of a fundamental modification of gravity. The prime example of such \textit{modified gravity} theories is the tensor-vector-scalar theory (TeVeS) of Bekenstein and Sanders \cite{Sa.97,Be.04,Sa.05}. Interesting connections between TeVeS and the class of Einstein-{\ae}ther theories \cite{JaMa.01} have been found \cite{Zl.al.07,Sk.08,Co.al.08}. Modified gravity theories are rather complicated extensions of general relativity (GR), and are for the moment not connected to fundamental physics. Moreover, they do not account for all the mass discrepancy at the intermediate scale of galaxy clusters \cite{An.al.08}. To resolve this difficulty a component of hot dark matter (HDM) in the form of massive neutrinos has been invoked \cite{An.al.07,An.09}. At cosmological scales the modified gravity theories also have some problems at reproducing the observed CMB spectrum \cite{Sk.al.06}, even when using a component of HDM.

The approach we propose below is able to successfully address both cosmological and galactic scales. We advocate that a non-standard form of dark matter may exist, while keeping the standard law of gravity (GR) unchanged. The physical belief of this alternative approach is the striking analogy between MOND and the electrostatics of (non-linear isotropic) dielectric media \cite{Bl1.07}. Indeed, the MOND equation \eqref{mond} can be interpreted as the standard Poisson equation if the gravitational field is sourced by baryonic matter \textit{and} by a ``digravitational'' medium playing the role of dark matter. The density of ``polarization masses'' in this medium is then $\rho_\text{pol} = - \bm{\nabla} \cdot \bm{\Pi}_\perp$ (anticipating the notation adopted below), where $\bm{\Pi}_\perp$ denotes the polarization field, which must be aligned with the local gravitational field,
\begin{equation}\label{Pig}
	\bm{\Pi}_\perp = - \frac{\chi(g)}{4\pi} \, \bm{g}\, .
\end{equation}
Here $\chi\equiv\mu-1$ denotes the ``gravitational susceptibility'' coefficient of the medium, while $\mu$ can be viewed as a ``digravitational'' constant. It was argued \cite{Bl1.07} that in the gravitational case the sign of $\chi$ should be negative, in agreement with what MOND predicts; indeed, we have $\mu<1$ in a straightforward interpolation between the MOND and Newtonian regimes, hence $\chi < 0$. Furthermore, arguments were given showing that the stability of the dipolar medium requires the existence of some environment-dependent internal non-gravitational force. More precisely, the force has to depend on the polarization field, i.e. the density of dipole moments.

Motivated by the previous interpretation of MOND we present in Section \ref{secII} an action principle for dark matter viewed as the gravitational analogue of a polarizable dielectric medium. In Section \ref{secIII} we show that this model is currently viable since it is in agreement with the standard cosmological scenario at large scales and recovers MOND at galactic scales. Some details regarding the variation of the action are relagated to Appendix \ref{appA}.

\section{Model of dipolar dark matter and dark energy}
\label{secII}

In previous work \cite{BlLe.08} (hereafter paper I; see also \cite{Bl2.07} for an earlier attempt) we proposed a relativistic model of dark matter and dark energy based on a particular concept of gravitational polarization. In contrast to modified gravity theories, the model should be viewed as a \textit{modified matter} theory. The idea that the phenomenology of MOND could arise from the CDM paradigm has been previously discussed \cite{KaTu.02,Mi2.02}. However here we shall consider a true modification of the physics of dark matter, drastically different from CDM (see also \cite{Br.al.09} for an alternative approach in a related spirit).

In paper I we showed that this particular model of modified dark matter permits recovering the phenomenology of MOND in a natural way, while being in agreement with the cosmological $\Lambda$-CDM model. The aim of this article is to prove that the model can be reformulated from a simple and physically meaningful matter action.

The dipolar medium is described as a fluid with mass current $J^\mu = \sigma u^\mu$, and endowed with a dipole moment vector $\xi^\mu$. Here $u^\mu=\ud x^\mu/\ud\tau$ is the time-like four-velocity of the fluid, with $\ud\tau = \sqrt{-g_{\mu\nu}\ud x^\mu\ud x^\nu}$ being the proper time (we pose $c=1$). The rest mass density reads $\sigma = \sqrt{-J_\mu J^\mu}$, and the mass current is conserved, i.e.
\begin{equation}\label{divJ}
	\nabla_\mu J^\mu = 0\, ,
\end{equation}
where $\nabla_\mu$ denotes the covariant derivative associated with the metric $g_{\mu\nu}$. The dipole moment $\xi^\mu$ has the dimension of a length, so that it is more like a displacement vector; the associated polarization field then reads $\Pi^\mu=\sigma\xi^\mu$. We have in mind that $\xi^\mu$ and $\Pi^\mu$ are effective variables resulting from an average performed at some macroscopic scale.

\begin{figure}
	\includegraphics[width=7cm]{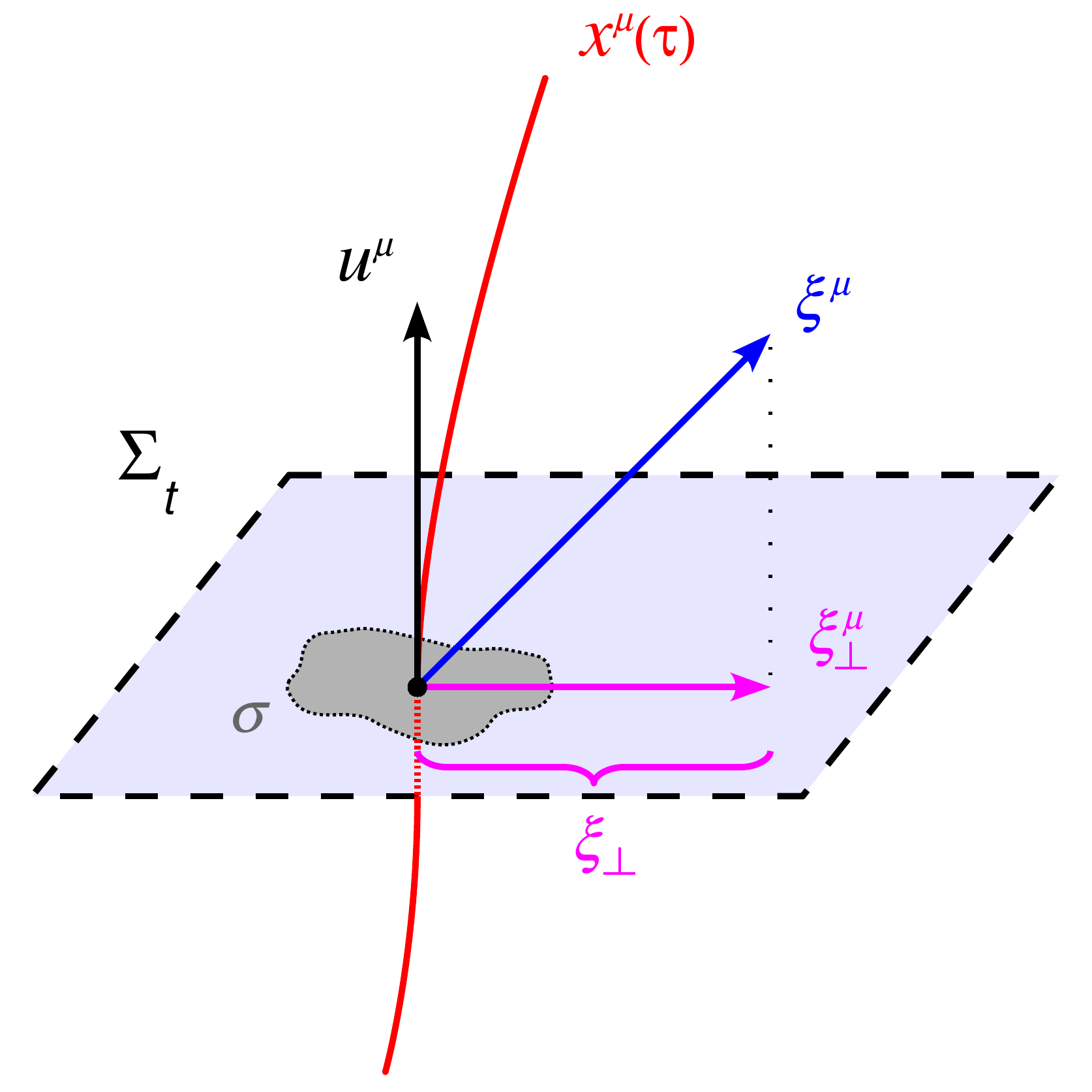}
	\caption{\footnotesize The dipolar fluid is entirely characterized by (i) its four-velocity $u^\mu$ tangent to the world-line $x^\mu(\tau)$, (ii) its rest mass density $\sigma$, and (iii) the dipole moment $\xi^\mu$. The polarization $\Pi_\perp = \sigma \xi_\perp$ is build from the norm $\xi_\perp$ of the projection $\xi_\perp^\mu$ of the dipole moment $\xi^\mu$ onto the space-like hypersurface $\Sigma_t$ orthogonal to $u^\mu$.}
	\label{fluid}
\end{figure}

The action of the dipolar dark matter is of the type $S = \int \ud^4 x \, \sqrt{-g} \, L$. It will be added to the Einstein-Hilbert action for gravity, and to the standard actions of all the other matter fields (baryons, photons, neutrinos, etc). We find that the Lagrangian consists of three terms: a mass term $\sigma$ in the ordinary sense (as for CDM), an interaction term built from the coupling between the current $J^\mu$ and the dipolar field $\xi^\mu$, and a potential scalar function $\calW$ characterizing an internal force acting on the dipolar particles, and depending on the polarization field. It explicitely reads
\begin{equation}\label{L}
	L = - \sigma + J_\mu \dot{\xi}^\mu - \calW (\Pi_\perp) \, .
\end{equation}
Both $J^\mu$ and $\xi^\mu$ will be considered as dynamical variables, to be varied independently when applying the principle of stationary action (cf. Appendix \ref{appA} for details). Here $\dot{\xi}^\mu \equiv \uD\xi^\mu/\ud\tau = u^\nu \nabla_\nu \xi^\mu$ denotes the proper time derivative of the dipole moment. Note that $\dot{\xi}^\mu$ is coupled to the current $J_\mu$ like an external field would be. However, here the dipole moment $\xi^\mu$ is an \textit{internal} field. The potential $\calW$ is assumed to depend solely on the norm $\Pi_\perp$ of the projection of the polarization field perpendicular to the four-velocity. Thus, $\Pi_\perp=\sigma\xi_\perp$ where $\xi_\perp = \sqrt{\perp_{\mu\nu} \! \xi^\mu \xi^\nu}$, with the usual orthogonal projector $\perp_{\mu\nu} \, = g_{\mu\nu} + u_\mu u_\nu$ (cf. Fig.~\ref{fluid}). As discussed in Sec.~\ref{secI}, the introduction of such an environment-dependent potential is motivated by the previous interpretation of the MOND phenomenology resulting from the mechanism of gravitational polarization.

A crucial point is that $\xi^\mu$ can be replaced in the coupling term of \eqref{L} by its orthogonal projection $\xi_\perp^\mu = \, \perp^\mu_\nu \! \xi^\nu$ without changing the dynamics. Indeed, a short calculation reveals that $J_\mu \dot{\xi}^\mu=J_\mu \dot{\xi}_\perp^\mu+\nabla_\mu(J^\mu u^\nu\xi_\nu)$, so that, because a pure divergence can be dropped from the Lagrangian, the only physical degrees of freedom are the three independent components of the vector $\xi_\perp^\mu$, which is \textit{space-like} (we denote $\dot{\xi}_\perp^\mu\equiv\uD\xi^\mu_\perp/\ud\tau$). This is to be contrasted with TeVeS and Einstein-{\ae}ther theories which are based on a fundamental \textit{time-like} vector field.

To obtain the equation of motion of the dipolar fluid we vary the action with respect to the dipole moment variable $\xi^\mu$, and get
\begin{equation}\label{motion}
	\dot{u}^\mu = - \mathcal{F}^\mu \equiv - \hat{\xi}_\perp^\mu \, \calW' \, ,
\end{equation}
where $\dot{u}^\mu\equiv\uD u^\mu/\ud\tau$ is the four-acceleration, $\hat{\xi}_\perp^\mu\equiv\xi_\perp^\mu/\xi_\perp$ is the unit direction along $\xi_\perp^\mu$, and $\calW' \equiv \ud \calW / \ud \Pi_\perp$. The motion is non-geodesic because of the internal force density $\mathcal{F}^\mu$ caused by the dipole moment $\xi^\mu$.

The variation with respect to $J^\mu$ yields the equation of evolution for the dipole moment. The constraint that the matter current is conserved, Eq.~\eqref{divJ}, is to be satisfied during the variation and we apply a convective variational procedure (see Appendix \ref{appA}). Defining for convenience the ``linear momentum'' $\Omega^\mu \equiv \dot{\xi}_\perp^\mu + u^\mu \left( 1 + 2 \xi_\perp \calW' \right)$, we obtain
\begin{equation}\label{evolution}
	\dot{\Omega}^\mu = \frac{1}{\sigma} \nabla^\mu \left( \calW - \Pi_\perp	\calW' \right) - \xi_\perp^\nu R^\mu_{\phantom{\mu} \rho \nu \lambda} u^\rho u^\lambda \, .
\end{equation}
This tells how the variation of the dipole moment should differ from parallel transport along the fluid's worldline. The first term on the right-hand-side (RHS) looks like a pressure term, while the second term represents the analogue of the standard coupling to Riemann curvature for spinning particles in GR \cite{Pa.51,BaIs.80}. Finally, varying with respect to the metric, we get the stress-energy tensor
\begin{equation}\label{Tmunu}
	T^{\mu \nu} = \Omega^{(\mu} J^{\nu)} - \nabla_\rho \left( \left[ \Pi_\perp^\rho u^{(\mu} - u^\rho \Pi_\perp^{(\mu} \right] u^{\nu)} \right) - g^{\mu\nu} \left( \calW - \Pi_\perp \calW' \right) .
\end{equation}
The RHS is made of a monopolar term associated with $\Omega^\mu$, while the second term is (minus) the divergence of a ``polarization'' tensor and is of a dipolar nature. Being proportional to the metric, the third term on the RHS will be related to a fluid of dark energy. We have $\nabla_\nu T^{\mu \nu}=0$ as a consequence of \eqref{motion}--\eqref{evolution}. We observe, in agreement with our earlier argument at the level of the Lagrangian, that all equations depend \textit{in fine} only on the perpendicular projection $\xi_\perp^\mu = \, \perp^\mu_\nu\xi^\nu$ of the dipole moment.

The equations of motion \eqref{motion} and evolution \eqref{evolution}, and the stress-energy tensor \eqref{Tmunu}, turn out to be exactly the same as in the model of paper I [see (2.20)--(2.21) and (2.24) there]. Those equations were derived starting from the more complicated Lagrangian given by (2.7) in paper I, and sharing some common features with the one for particles with spin moving in an arbitrary background \cite{BaIs.80}. Furthermore, they were obtained after imposing a particular choice of solution satisfying some consequence of the initial equations ($\Xi = 1$ in the notation of paper I). Despite this rather complicated way to derive them, it was found that the equations provide the sensible physics for a successful model of dark matter and dark energy. We have now proved that the same equations derive directly (without any further assumptions) from the remarkably simple Lagrangian \eqref{L}, which lends itself better to physical interpretation.

\section{Recovering the standard cosmological model and MOND}
\label{secIII}

\begin{figure}
	\includegraphics[width=9cm]{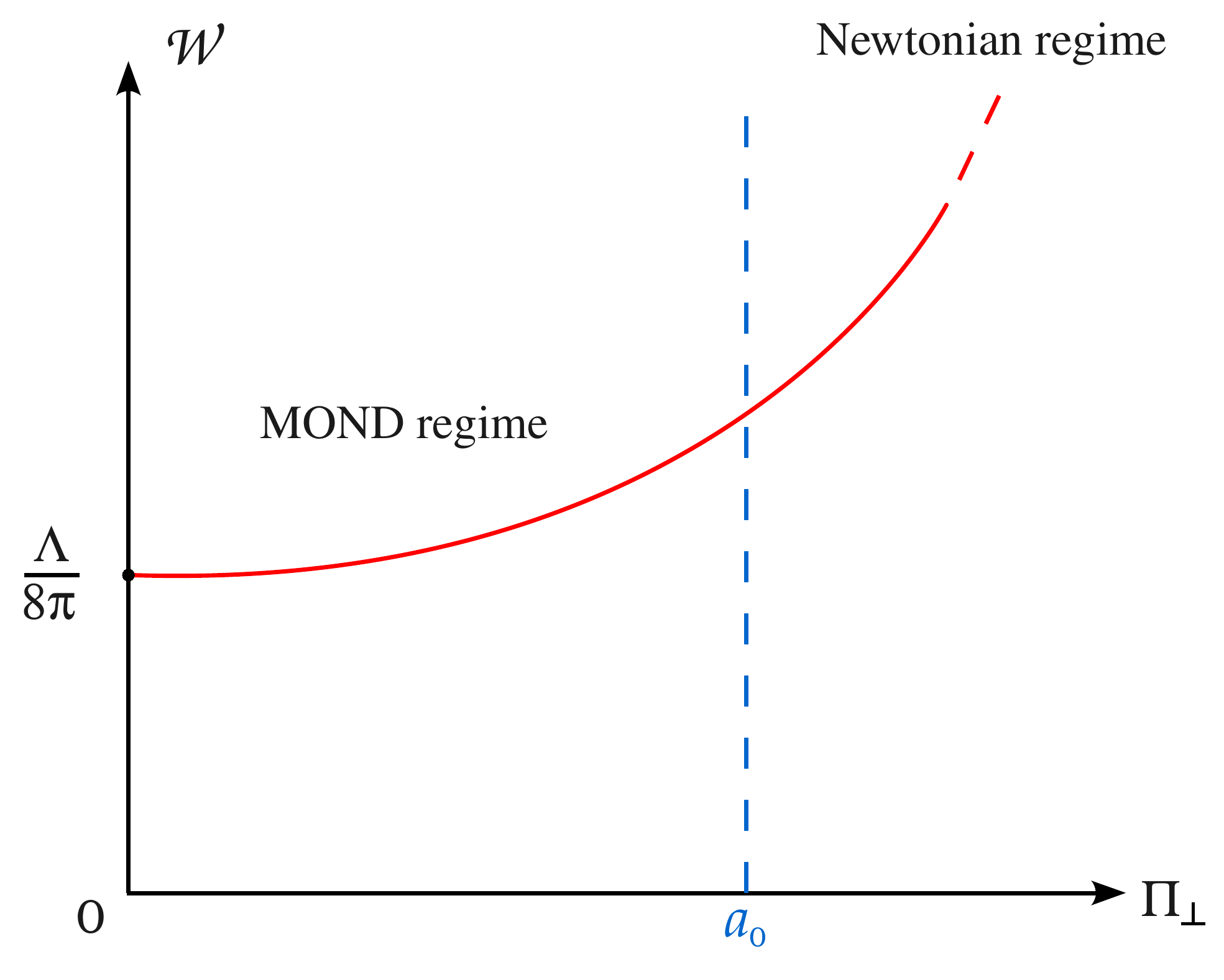}
	\caption{\footnotesize The potential $\calW$ is function of the polarization field $\Pi_\perp$. It has an anharmonic behavior in the MOND regime $\Pi_\perp \ll a_0$, and its minimum is given by the cosmological constant $\Lambda$. The leading order non-harmonicity is parametrized by the MOND acceleration scale $a_0$ [cf. Eq.~\eqref{W}]. The Newtonian regime $\Pi_\perp \gg a_0$ is discussed in details in paper I.}
	\label{potential}
\end{figure}

We now review the main consequences of this model, presenting alternative versions of most arguments compared to paper I. To achieve agreement with MOND and with $\Lambda$-CDM (to first-order cosmological perturbations), we have to fine-tune the potential $\calW$ in the action. Indeed, we find that $\calW$ is ``phenomenologically'' determined up to third order in an expansion when the polarization field $\Pi_\perp$ tends to zero. Physically, this corresponds to $\Pi_\perp\ll a_0$, which in turn will mean that gravity is weak, $g\ll a_0$, like in the outskirts of a galaxy or in a nearly homogeneous and isotropic cosmology. In this regime $\calW$ takes the anharmonic form
\begin{equation}\label{W}
	\calW(\Pi_\perp) = \frac{\Lambda}{8 \pi} + 2 \pi \, \Pi_\perp^2 + \frac{16 \pi^2}{3 a_0} \, \Pi_\perp^3 + \mathcal{O}(\Pi_\perp^4) \, .
\end{equation}
The minimum is directly related to the cosmological constant $\Lambda$, and the deviations from that minimum are fixed by the agreement with MOND; in particular $a_0$ parametrizes the third-order deviation (see Fig.~\ref{potential}).

Let us assume, following paper I, that the theory depends only on one new fundamental scale --- the constant MOND acceleration $a_0$. When entering the MOND regime, $\Pi_\perp/a_0$ is of order one, therefore $\calW$ naturally scales with $a_0^2$. If $\calW$ is to come from some fundamental theory, we expect that the dimensionless coefficients in the expansion \eqref{W} after global rescaling by $a_0^2$ should be of the order of one. In particular, $\Lambda$ should itself be of the order of $a_0^2$. As is well known \cite{Mi.02}, the current astrophysical measurements verify the ``cosmic coincidence'' that $\Lambda\sim a_0^2$. This is a natural consequence of our model.

\subsection{First-order cosmological perturbations}

We now turn to the application at early cosmological time, where we consider a linear perturbation around an homogeneous and isotropic Friedman-Lema\^itre-Robertson-Walker (FLRW) universe. Since the dipole moment $\xi_\perp^\mu$ is space-like, it will break the spatial isotropy of the FLRW background, and must necessarily belong to the first-order perturbation, which we indicate by $\xi_\perp^\mu=\mathcal{O}(1)$. For instance, from \eqref{W} we find that the internal force is also of first order, $\mathcal{F}^\mu = 4\pi\,\Pi_\perp^\mu+\mathcal{O}(2)$. At that order the stress-energy tensor \eqref{Tmunu} simplifies very much, and can be decomposed into dark energy and dark matter components, namely $T^{\mu \nu} = T_\text{de}^{\mu \nu} + T_\text{dm}^{\mu \nu}$, where the dark energy is simply given by the cosmological constant, $T_\text{de}^{\mu \nu} = - \frac{\Lambda}{8 \pi} \, g^{\mu \nu}+\mathcal{O}(2)$, while the dark matter reads
\begin{equation}\label{Tmunucosm2}
	T_\text{dm}^{\mu \nu} = \rho\,\tilde{u}^\mu \tilde{u}^\nu+\mathcal{O}(2)\, .
\end{equation}
Here $\rho\equiv\sigma-\nabla_\mu\Pi^\mu_\perp$ is the energy density of the dark matter fluid, and $\tilde{u}^\mu = u^\mu + \dot{\xi}^\mu_\perp - \xi^\nu_\perp \nabla_\nu u^\mu$ (i.e. $\tilde{u}^\mu = u^\mu - \mathscr{L}_{\xi_\perp}u^\mu$, where $\mathscr{L}_{\xi_\perp}$ is the Lie derivative) is an effective four-velocity field, which satisfies $\tilde{u}_\mu\tilde{u}^\mu=-1+\mathcal{O}(2)$ and the approximate conservation law $\nabla_\mu(\rho\,\tilde{u}^\mu)=\mathcal{O}(2)$. This shows that, at linear order, the dark matter cannot be distinguished from a pressureless perfect fluid; in particular the fluid's motion is geodesic, $\tilde{u}^\nu \nabla_\nu \tilde{u}^\mu=\mathcal{O}(2)$. Therefore, the model makes the same predictions as the $\Lambda$-CDM cosmological model at linear order (see paper I for more details). In particular, adjusting the background value of $\rho$ (namely $\bar{\rho}$ such that $\rho=\bar{\rho}+\mathcal{O}(1)$; notice that $\bar{\rho}=\bar{\sigma}$) to the measured value of dark matter today, $\Omega_\text{dm} \simeq 0.23$, and choosing $\Lambda$ in such a way that the dark energy contribution is $\Omega_\text{de} \simeq 0.73$, we are in agreement with the observed fluctuations of the CMB. To be more precise, the linearized perturbation equations, given by (3.48)--(3.49) in paper I, are identical with those of $\Lambda$-CDM with no additional degrees of freedom, since the dipole moment has been absorbed at linear order into the effective vector field $\tilde{u}^\mu$ and mass density $\rho$. Therefore the model reproduces both the location and the height of the peaks of the CMB.

At non-linear order in cosmological perturbations, the model should differ from the standard $\Lambda$-CDM scenario. The fluid's dynamics will no longer be geodesic. We expect that the formation of large scale structures will be triggered not by the monopolar part $\sigma$ of dark matter, which should not cluster much (see below), but by the dipolar component present in $\rho = \sigma - \nabla_\mu \Pi^\mu_\perp$. This should be checked performing numerical simulations in cosmology.

\subsection{Non-relativistic limit}

Next we investigate the dipolar medium around a typical galaxy at low redshift. For this purpose it is sufficient to consider the non-relativistic (NR) limit of the model, when $c\rightarrow +\infty$. Working at the level of the Lagrangian \eqref{L}, we find that $\mathcal{L}=\sqrt{-g} \, L$ reduces in this limit to
\begin{equation}\label{L_NR}
	\mathcal{L}_\text{NR} = \sigma_\star \!\left( \frac{\bm{v}^2}{2} + U + \bm{g} \cdot \bm{\xi}_\perp + \bm{v} \cdot \frac{\ud \bm{\xi}_\perp}{\ud t} \! \right) - \calW (\Pi_\perp) \, .
\end{equation}
Here $\bm{v}$ is the fluid's ordinary velocity, $\bm{\xi}_\perp$ is the dipole moment vector, and $\bm{g} = \bm{\nabla} U$ is the Newtonian gravitational field with $U$ the gravitational potential. The rest mass density $\sigma_\star$ satisfies the usual continuity equation $\partial_t \sigma_\star + \bm{\nabla} \cdot \bm{J}_\star= 0$, where the current reads $\bm{J}_\star=\sigma_\star\bm{v}$. We denote by $\ud/\ud t=\partial_t+\bm{v}\cdot\bm{\nabla}$ the convective time derivative, so that e.g. $\ud\sigma_\star/\ud t=-\sigma_\star\bm{\nabla}\cdot\bm{v}$. We discarded for convenience the rest mass term ($-\sigma_\star$) in the NR Lagrangian \eqref{L_NR}. From the coupling term $J_\mu \dot{\xi}^\mu$ in the Lagrangian \eqref{L}, we recover in the NR limit \eqref{L_NR} the gravitational analogue $\bm{g} \cdot \bm{\Pi}_\perp $ of the coupling of the polarization field to an external electric field. Notice also the extra term $\bm{J}_\star \cdot \ud \bm{\xi}_\perp / \ud t$ in \eqref{L_NR}, which arises in the gravitational case.

The NR equations of motion and evolution are obtained by varying the Lagrangian \eqref{L_NR} with respect to $\bm{\xi}$ and $\xi^0$ (note that $\bm{\xi}_\perp=\bm{\xi}-\bm{v}\,\xi^0$), and $\bm{J}_\star$. We get
\begin{align}\label{eomNR}
	\frac{\ud \bm{v}}{\ud t} &= \bm{g} - \bm{\mathcal{F}} \, , \\ \label{evolNR}
	\frac{\ud^2 \bm{\xi}_\perp}{\ud t^2} &= \bm{\mathcal{F}} + \frac{1}{\sigma_\star} \bm{\nabla} \left( \calW - \Pi_\perp \, \calW' \right) + \left( \bm{\xi}_\perp\!\cdot \!\bm{\nabla} \right) \bm{g} \, ,
\end{align}
in agreement with the NR limits of \eqref{motion} and \eqref{evolution}. The gravitational equation follows from adding to \eqref{L_NR} the Newtonian Lagrangian $\mathcal{L}_U=-\frac{1}{8\pi}\bm{\nabla}U\!\cdot\!\bm{\nabla}U$ (coming from the NR limit of the Einstein-Hilbert action in GR) and the contribution of baryons. Varying with respect to $U$ gives
\begin{equation}\label{poisson}
	\bm{\nabla} \! \cdot \left( \bm{g} -4 \pi \, \bm{\Pi}_\perp \right) = -4 \pi \left( \rho_\text{b} + \sigma_\star \right) .
\end{equation}

We proposed in paper I a mechanism by which the dipolar medium does not cluster as much as baryonic matter during the cosmological evolution. This is supported by an exact solution of \eqref{eomNR}--\eqref{poisson}, valid in spherical symmetry, where the dipolar fluid has zero velocity, $\bm{v}=\bm{0}$, and a constant mass density $\sigma_\star$ (see Appendix A in paper I). The dipole moments remain at rest because the gravitational field $\bm{g}$ is balanced by the internal force $\bm{\mathcal{F}}$. From this we inferred the hypothesis of ``weak-clustering'', namely that the typical mass density of dipole moments in a galaxy (after cosmological evolution) is much less than the baryonic density, $\sigma_\star\ll\rho_\text{b}$, and perhaps of the order of the mean cosmological value, $\sigma_\star \sim \bar{\sigma}_\star$. Furthermore the dipolar medium is essentially static, $\bm{v}\simeq\bm{0}$. If this hypothesis is true, we have $\bm{g}\simeq\bm{\mathcal{F}}$ by \eqref{eomNR}, so the polarization field $\bm{\Pi}_\perp$ is aligned with the gravitational field $\bm{g}$, i.e. the medium is polarized. Using $\bm{\mathcal{F}}=\hat{\bm{\xi}}_\perp\,\calW'$ together with the expression
of the potential \eqref{W}, we get
\begin{equation}\label{gPi}
	\bm{g} \simeq 4\pi\,\bm{\Pi}_\perp\biggl(1+4\pi\frac{\Pi_\perp}{a_0}\biggr) + \mathcal{O}(\Pi_\perp^3) \, .
\end{equation}
Hence the gravitational susceptibility coefficient $\chi=\mu-1$ defined by \eqref{Pig} takes the appropriate form in the MOND regime, namely $\chi(g)\simeq -1 + g / a_0 + \mathcal{O}(g^2)$. We conclude that \eqref{poisson} is equivalent to the MOND equation \eqref{mond}. (See paper I for a discussion of the Newtonian regime $g \gg a_0$.) Note that it is crucial that we could neglect the monopolar part $\sigma_\star$ of the dipolar medium as compared to $\rho_\text{b}$, so that galaxies appear baryonic in MOND fits of the rotation curves. On the other hand, the monopolar dark matter $\sigma_\star$ as we have seen plays the dominant role in a cosmological context. It may also help explaining the missing dark matter at the intermediate scale of galaxy clusters \cite{An.al.08}.

The weak-clustering mechanism also tells us that the evolution of the dipole moments should be slow. In spherical symmetry, the two last terms of \eqref{evolNR} cancel each other, and we get $\partial_t^2\Pi_\perp = 4 \pi \sigma_\star \Pi_\perp$ in the MOND regime. This shows the presence of an instability, with exponentially growing modes. However the unstable modes will develop on the self-gravitating time scale $\tau_\text{g} = \sqrt{\pi / \sigma_\star}$, which is very long thanks to $\sigma_\star\ll\rho_\text{b}$. Using the mean cosmological value $\bar{\sigma}_\star \simeq 10^{-26}~\text{kg} / \text{m}^3$ we get $\tau_\text{g} \simeq 6 \times 10^{10}~\text{years}$. Thus this instability is not a problem classically.

\section{Conclusion}

In conclusion, the model (i) explains the phenomenology of MOND by the physical process of gravitational polarization, (ii) makes a unification between the dark matter \textit{\`a la MOND} and the dark energy in the form of a cosmological constant (with the interesting outcome that $\Lambda\sim a_0^2$), and (iii) recovers the successful standard cosmological model $\Lambda$-CDM at linear perturbation order. However the model lacks some connection to microscopic physics and describes the dipole moments in an effective way; notably the potential $\calW$ in \eqref{W} is for the moment purely phenomenological. The model should be further tested in cosmology, by studying second-order cosmological perturbations where we expect a departure from $\Lambda$-CDM, by computing numerically the non-linear growth of perturbations and formation of large scale structures, and by investigating the intermediate scale of galaxy clusters.

\appendix

\section{Variation of the action functional}\label{appA}

Here we provide some details on the derivation of the equations of motion and evolution of the dipolar fluid. They derive from an action of the general form
\begin{equation}\label{action}
	S = \int \ud^4 x \, \sqrt{-g} \, L[J^\mu,\xi^\mu,g_{\mu\nu}]\,,
\end{equation}
where as indicated the Lagrangian density $L$ is a functional of the matter current $J^\mu$, the dipole moment $\xi^\mu$ (and its covariant derivative $\nabla_\nu \xi^\mu$), and the covariant metric $g_{\mu\nu}$. 

We vary first the action with respect to the dipole moment $\xi^\mu$. Notice that in our Lagrangian \eqref{L} the dependence on $\nabla_\nu \xi^\mu$ is only through the covariant \textit{time} derivative $\dot{\xi}^\mu$. In that case, denoting the conjugate momentum of the dipole by $\Psi_\mu\equiv\partial L/\partial \dot{\xi}^\mu$, we obtain from the principle of stationary action
\begin{equation}\label{var1}
	 \dot{\Psi}_\mu + \Theta \,\Psi_\mu = \frac{\partial L}{\partial \xi^\mu} \, ,
\end{equation}
with $\Theta\equiv\nabla_\nu u^\nu$. Since the vector field $\xi^\mu$ is unconstrained, this equation is equivalent to the standard Lagrange equation
\begin{equation}
	 \nabla_\nu \biggl( \frac{\partial L}{\partial \nabla_\nu \xi^\mu} \biggr) = \frac{\partial L}{\partial \xi^\mu} \, .
\end{equation}
In the case at hands of the Lagrangian \eqref{L} we then obtain the equation of motion of the dipolar fluid as given by \eqref{motion}. 

However, the variation with respect to the current $J^\mu$ is trickier because of the constraint that this current is conserved: $\nabla_\mu J^\mu=0$. We adopt a convective variational approach \cite{Ta.54,Ca.91,CaKh.92} in which the variation $\delta J^\mu$ is constrained to have the form which is precisely induced by an infinitesimal displacement of the flow lines of $J^\mu$. Denoting $\delta x^\mu$ the generator of the displacement of the flow lines we have
\begin{equation}\label{deltaJ}
	\delta J^\mu = \delta x^\nu \nabla_\nu J^\mu - J^\nu \nabla_\nu \delta x^\mu + J^\mu \nabla_\nu \delta x^\nu\,,
\end{equation}
which is automatically divergenceless: $\nabla_\mu \delta J^\mu = 0$. The variation of the action with respect to $J^\mu$, using the fact that $\delta x^\mu$ is unconstrained, then yields
\begin{equation}\label{var2}
	J^\nu \bigl[ \nabla_\nu p_\mu - \nabla_\mu p_\nu \bigr] = 0 \, ,
\end{equation}
with $p_\mu\equiv\partial L/\partial J^\mu$ being the momentum associated with the current. In the case of a perfect fluid this equation is equivalent to the usual Euler equation, where $p_\mu$ in that case is the current of enthalpy \cite{Sy.38,Lic}. For the Lagrangian \eqref{L}, the equation \eqref{var2} yields the equation of evolution of the dipolar fluid in the form \eqref{evolution}.

Finally the stress-energy tensor is derived by variation of the Lagrangian with respect to the metric. We take into account the dependence of the current $J^\mu$ on the metric through the volume element $\sqrt{-g}\,\ud^4 x$, which means that the so-called ``coordinate'' current density $J_*^\mu=\sqrt{-g}J^\mu$ is the relevant metric-independent variable. In addition we treat the change in the metric that is hidden into the covariant time derivative $\dot{\xi}^\mu$ by means of the Palatini formula. The result is
\begin{equation}\label{Tmunugen}
	T^{\mu \nu} = 2 \frac{\partial L}{\partial g_{\mu \nu}} + g^{\mu \nu} \Bigl( L - J^\rho p_\rho \Bigr) + u^\mu u^\nu \, \dot{\xi}^\rho \Psi_\rho + \nabla_\rho \Bigl( u^{(\mu} \xi^{\nu)} \Psi^\rho - u^\rho \xi^{(\mu} \Psi^{\nu)} - \xi^\rho u^{(\mu} \Psi^{\nu)} \Bigr) \, .
\end{equation}
(Notice the misprint in the first dipolar term in the corresponding equation (2.22) of \cite{BlLe.08}.) Straightforward calculations in the case of the Lagrangian density $\eqref{L}$ give the explicit expression \eqref{Tmunu}.

\bibliography{/tmp_mnt/netapp/users_home4/letiec/Articles/ListeRef}

\end{document}